\documentclass[aps,prb,twocolumn]{revtex4}
\usepackage{amsmath}
\usepackage{graphicx}
\usepackage{multirow}
\usepackage{array}
\usepackage{inputenc}
\usepackage{epsfig}
\usepackage{graphicx}
\usepackage{inputenc}
\usepackage{bm}

\newcommand{\dyti}{Dy$_2$Ti$_2$O$_7$}
\newcommand{\tbti}{Tb$_2$Ti$_2$O$_7$}
\newcommand{\tbsn}{Tb$_2$Sn$_2$O$_7$}
\newcommand{\gdti}{Gd$_2$Ti$_2$O$_7$}

\newcommand{\ybti}{Yb$_2$Ti$_2$O$_7$}
\newcommand{\rti}{R$_2$Ti$_2$O$_7$}
\newcommand{\mub}{$\mu_{\rm B}$}

\newcommand{\R}{R$^{3+}$}
\begin{document}
\author{S. Guitteny$^{1}$, I. Mirebeau$^{1}$, P. Dalmas de R\'eotier$^{2,3}$, C. V. Colin$^{4,5}$, P. Bonville$^{6}$, F. Porcher$^{1}$, B. Grenier$^{2,3}$, C. Decorse$^{7}$ and S. Petit$^{1}$}
\affiliation{$^1$ CEA, Centre de Saclay, DSM/IRAMIS/ Laboratoire L\'eon Brillouin, F-91191 Gif-sur-Yvette, France}
\affiliation{$^2$ Universit\'e Grenoble Alpes, INAC-SPSMS, 38000 Grenoble, France}
\affiliation{$^3$ CEA, INAC-SPSMS, 38000 Grenoble, France}
\affiliation{$^4$ Univ. Grenoble Alpes, Inst NEEL F-38000 Grenoble, France}
\affiliation{$^5$ CNRS, Inst NEEL, F-38000 Grenoble, France}
\affiliation{$^6$ CEA, Centre de Saclay, DSM/IRAMIS/ Service de Physique de l'Etat Condens\'e, F-91191 Gif-Sur-Yvette, France}
\affiliation{$^7$ ICMMO, Universit\'e Paris-Sud, F-91400 Orsay France}
\title{Mesoscopic correlations in \tbti\, spin liquid}
\date{\today}
\begin{abstract}
We have studied the spin correlations with $\bf{k}$= ($\frac12$, $\frac12$, $\frac12$) propagation vector which appear below 0.4\, K in \tbti\ spin liquid by combining powder neutron diffraction and specific heat on Tb$_{2+x}$Ti$_{2-x}$O$_{7+y}$ samples with $x$=0, 0.01, -0.01. The $\bf{k}$= ($\frac12$, $\frac12$, $\frac12$) order clearly appears on all neutron patterns by subtracting a pattern at 1.2(1)\,K. Refining the subtracted patterns at 0.07\,K yields two possible spin structures, with spin-ice-like and monopole-like correlations respectively. Mesoscopic correlations involve Tb moments of 1 to 2 \mub\ ordered on a length scale of about 20 \AA. In addition, long range order involving a small spin component of 0.1 to 0.2 \mub\ is detected for the $x$= 0 and 0.01 samples showing a peak in the specific heat. Comparison with previous single crystals data suggests that the ($\frac12$, $\frac12$, $\frac12$) order settles in through nanometric spin textures with dominant spin ice character and correlated orientations, analogous to nanomagnetic twins. 
\end{abstract}
\pacs{81.05.Bx,81.30.Hd,81.30.Bx, 28.20.Cz}
\maketitle
Rare earth pyrochlores with geometrical magnetic frustration are model compounds to investigate classical spin ice as well as quantum spin ice physics. The short range ordered ground states with large degeneracy give rise to a large variety of interesting phenomena such as magnetic monopoles\cite{Castelnovo2008,Ryzhkin2011}, first order transition \cite{Chang2012}, or unconventional magnetoelastic excitations \cite{Guitteny2013,Fennell2014}. These frustrated ground states are highly sensitive to small perturbations and to the presence of defects. Already in the 80s, Villain \cite{Villain1979} predicted that for an arbitrary small concentration of defects the spinel lattice with pyrochlore structure could switch its ground state from spin liquid to spin glass. More recently many studies 
focus on the role of a minute amount of defects in pyrochlores. The crystal structure of \rti\ pyrochlores (where \R\ is a rare earth ion, Ti$^{4+}$ is non magnetic) accommodates oxygen vacancies, small off-stoechiometries in the \R or Ti$^{4+}$ lattice, "stuffing" or site inversion. Oxygen vacancies introduce Ti$^{3+}$ defects instead of Ti$^{4+}$ ions, modifying the R$^{3+}$ crystal field and inducing local distortions, with various consequences on the low temperature magnetic properties and spin ice physics. In classical spin ice \dyti, oxygen depletion or a low level Dy stuffing in the Ti sites strongly slows down the monopole dynamics at very low temperature \cite{Revell2013,Sala2014} and affects the non-equilibrium macroscopic properties \cite{Pomaranski2013,Paulsen2014}. 
In the quantum spin ice \ybti, small off-stoechiometry clearly affects the magnetic ground state \cite{Dortenzio2013}. It appears that precise studies of the magnetic defects are needed to resolve controversies concerning these unconventional ground states.

Among the pyrochlores, \tbti\ is one of the most complex, the subject of numerous investigations since the discovery of its spin liquid character in 1999. The Tb$^{3+}$
short range correlated moments fluctuate \cite{Gardner1999} down to 20\,mK at the time scale of the muon probe (10$^{-6}$s). The origin of these fluctuations has been highly debated \cite{Molavian2007,Bonville2011,Gaulin2011,Petit2012,Sazonov2013,Curnoe2013}. Several mechanisms have been proposed taking into account the non-Kramers nature of the Tb ion and the small energy gap between the ground and first excited crystal field doublets, which open the possibility of quantum spin ice behavior \cite{Gingras2014}. Recent studies of the crystal and magnetic excitations using polarized neutrons show that the magneto-elastic coupling plays a crucial role, yielding an interaction between the first excited crystal field level and an acoustic phonon branch, resulting in a hybrid phonon-magnon mode \cite{Guitteny2013,Fennell2014}. The magnetoelastic coupling also induces original properties in the pyrochlore family, such as giant magnetostriction \cite{Mamsurova1986,Klekovina2011} and very low thermal conductivity \cite{Li2013}. All these features suggest that the \tbti\, ground state is determined by an interaction between quadrupolar moments, for which a dynamical Jahn-Teller distortion could be the first (mean-field) approximation \cite{Sazonov2013,Bonville2014,Gehring1975}. 

 In \tbti\ an exotic transition has been observed around 0.4\,K, as shown initially by anomalies of the specific heat \cite{Yaouanc2011}. Below this temperature, spin fluctuations coexist with static spin correlations and spin glass-like irreversibilities \cite{Lhotel2012,Legl2012}.
 In single crystals, these correlations have been detected by an energy analysis of the neutron cross section \cite{Petit2012,Fennell2012,Fritsch2013,Fritsch2014}, yielding an elastic contribution with diffuse maxima at the positions of half integer Bragg peaks. This intensity vanishes under a small magnetic field of about 0.1\,T and shows some sample dependence, suggesting the influence of a minute amount of defects. 
 In powders, a systematic study of Tb$_{2+x}$Ti$_{2-x}$O$_{7+y}$ samples \cite{Taniguchi2013} showed that the system could be tuned by minute changes of the Tb content around a critical concentration $x_c$=-0.005 from a purely spin liquid state for $x$ $\leq$ $x_c$ to a state with partial antiferromagnetic order with $\bf{k}$= ($\frac12$, $\frac12$, $\frac12$) propagation vector for $x$ $>$ $x_c$. The detailed nature of this $\bf{k}$= ($\frac12$, $\frac12$, $\frac12$) order was not determined 
  in  Ref. \onlinecite{Taniguchi2013}, but it seemed to extend over mesoscopic length scales, rather than the infinite length scale associated with symmetry breaking and Bragg scattering. 

 To investigate this magnetic order, we studied powder samples of Tb$_{2+x}$Ti$_{2-x}$O$_{7+y}$ with a precise control of the stoechiometry $x$= -0.01, 0 and 0.01, by combining neutron diffraction and specific heat measurements. 
    In all samples, by subtracting a pattern at 1.2(1)\,K, we observe magnetic peaks of Lorentzian line shape, which can be indexed with the $\bf{k}$= ($\frac12$, $\frac12$, $\frac12$) propagation vector. These peaks correspond to the mesoscopic order discussed in the following. Its length scale deduced from the Lorentzian line shape is around 20 \AA, which corresponds to one magnetic cubic unit cell. Moreover,
 in the samples ($x$=0, 0.01) showing a transition in the specific heat, we also observe a Bragg-like contribution, involving a minute Tb moment ordered at a length scale of about 500 \AA, close to the resolution limit. Both Bragg-like and Lorentzian contributions vanish around 0.4\,K. By refining the magnetic patterns at the lowest temperature (0.07\,K), we find two possible spin arrangements, 
 fitting the data equally well. In both cases, the Tb moments remain aligned along their local $\langle$111$\rangle$ axes, therefore respect the Ising character of the crystal field. The first arrangement corresponds to magnetic moments in an ordered spin ice configuration \cite{Mirebeau2005}, antiferromagnetically coupled along the $\langle$100$\rangle$ cubic axes \cite{Fritsch2013} without violating the ice rules. The second one is described by a piling along a $[$111$]$ axis of monopole and antimonopole layers \cite{Sazonov2012} separated by spin ice layers. Comparing the calculated single crystal maps to the experimental one suggests that spin ice-like configurations with non-random mutual orientations, describing a $"$nanomagnetic twin texture$"$, are actually favored at 
 $T=0$. The monopole-like structure could be also stabilized when heating. We discuss the relationship between the mesoscopic order detected in the powder data and the diffuse scattering features (anisotropic excitations, pinch points) shown by single crystals. 
 \begin{figure}[ht]
\centerline{\includegraphics[width=8cm]{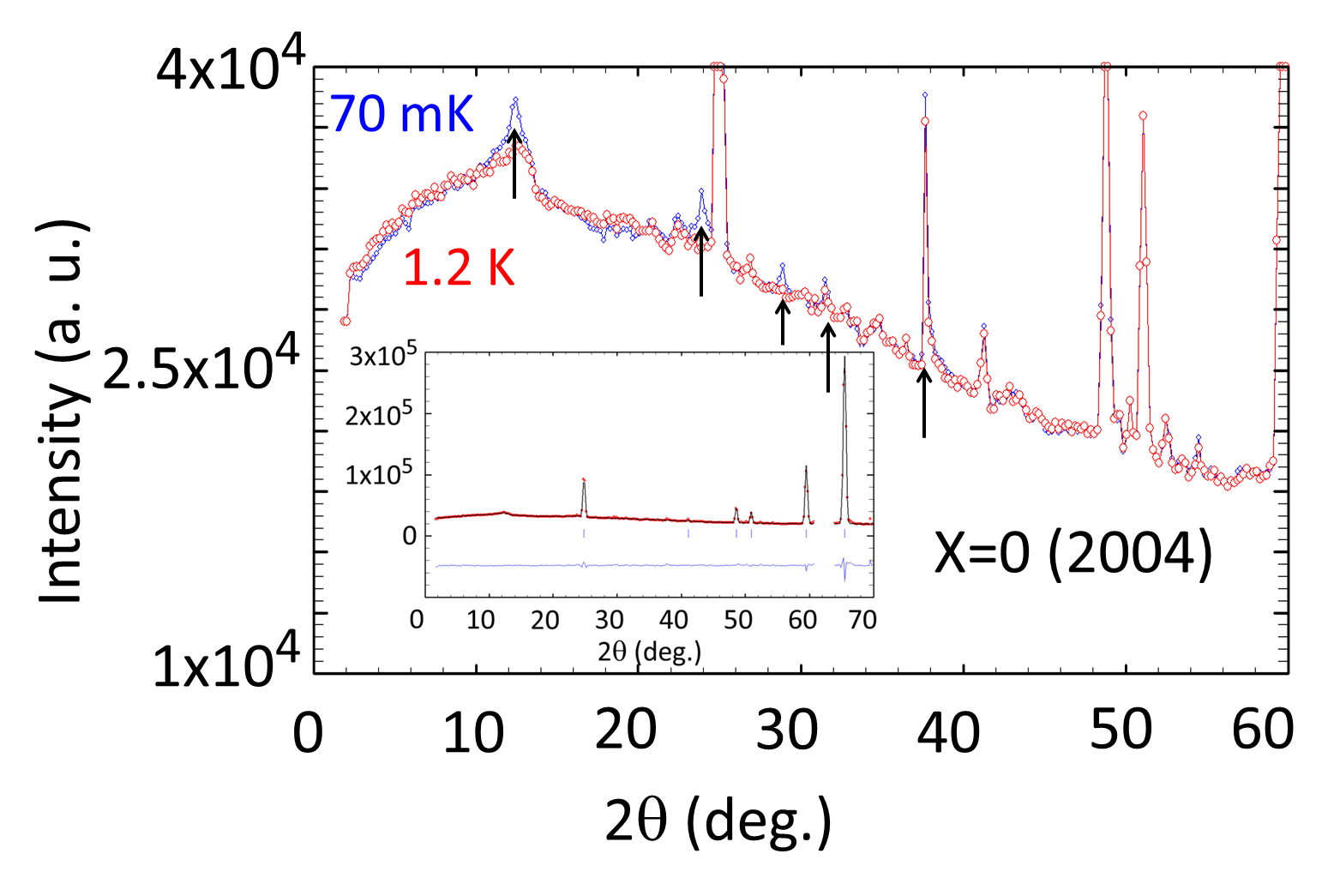}}
\caption{
(color on line) Tb$_2$Ti$_2$O$_7$. Neutron diffraction patterns measured at 0.07\,K and 1.2\,K, focusing on the region of the magnetic peaks, as shown by arrows. The full pattern at 0.07\,K is shown in the inset,  with the refinement of the crystal structure yielding the scaling of the ordered magnetic moment.
}
\label{figunsubtracted}
\end{figure}
 \begin{figure}[ht]
\centerline{\includegraphics[width=8cm]{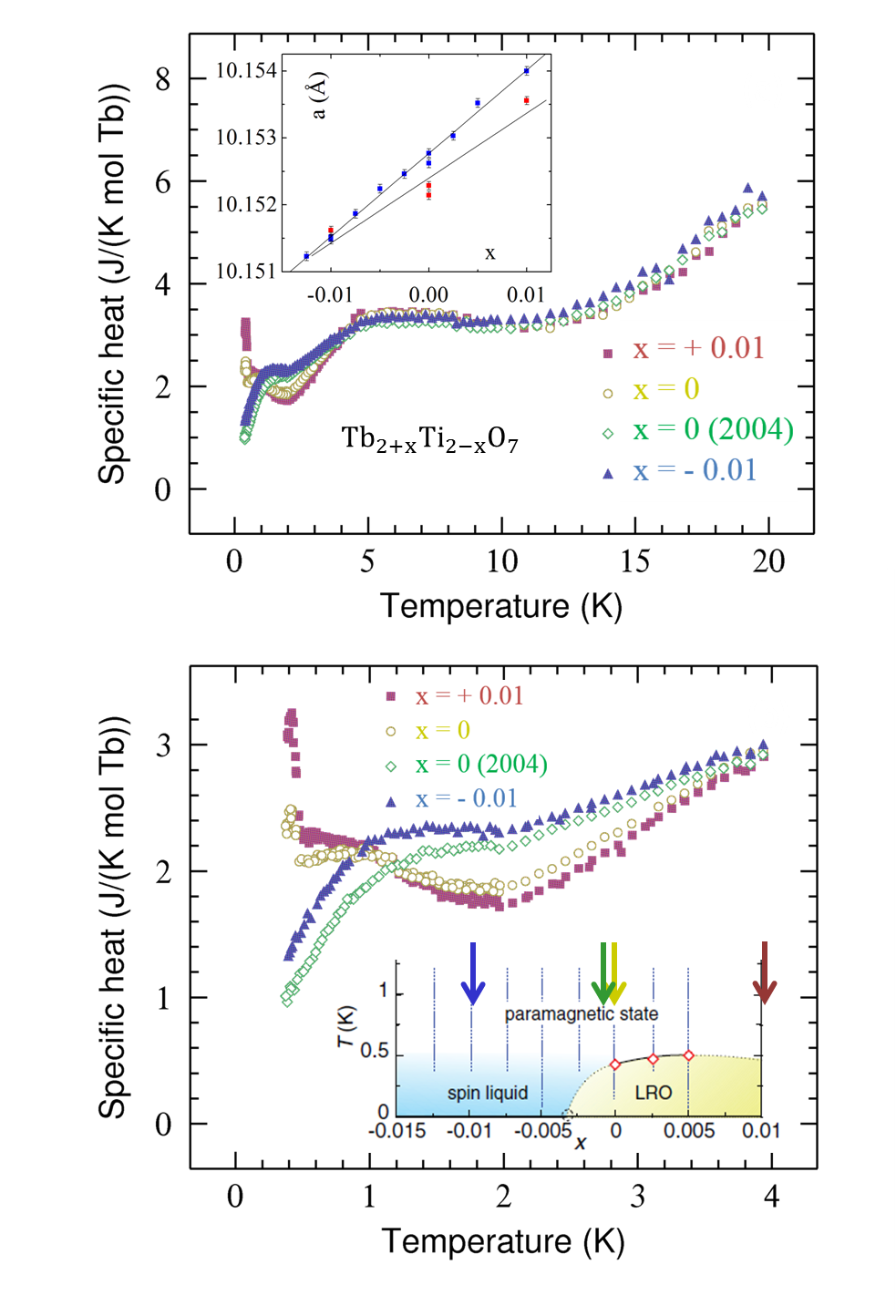}}
\caption{
(color on line) Tb$_{2+x}$Ti$_{2-x}$O$_{7+y}$. (a) Total specific heat $C_{\rm p}(T)$ versus temperature $T$ for several $x$ contents. In inset the lattice constant of the cubic cell $\it{a}$ deduced from X-ray data at 300\,K (red dots), in comparison with values from Ref. \onlinecite{Taniguchi2013} revised in 2015 (blue dots). (b) $C_{\rm p}(T)$  curves focusing on the low temperature region. In inset, the phase diagram of Ref. \onlinecite{Taniguchi2013} where the arrows show the sample concentrations studied here.
}
\label{figcp}
\end{figure} 

\section{Experiment}
Powder samples of Tb$_{2+x}$Ti$_{2-x}$O$_{7+y}$ with $x$= 0, 0.01 and -0.01 were synthesized as described in Ref. \onlinecite{Taniguchi2013}. The value of $x$ was adjusted by changing the mass of the starting constituents, Tb$_4$O$_7$ and TiO$_2$, which were heated in air at 1350$^{\circ}$C for several days with periodic grindings to ensure complete reactions. The obtained samples were ground into powder then annealed at 800$^{\circ}$C in air for one day. Another stoechiometric sample, synthesized in 2004 using a similar procedure \cite{Gardner1999} and previously studied with neutrons \cite{Mirebeau2002,Mirebeau2007} was also measured for comparison. 

All samples were characterized by X-ray diffraction and high resolution neutron diffraction patterns were also recorded for the stoechiometric samples on the spectrometer 3T2 at the Laboratoire L\'eon Brillouin (LLB), yielding a crystal structure in agreement with previous results. Slight colour changes between the samples with $x$=-0.01 (beige), 0 (white), +0.01 (light brown) are attributed to small variations in the concentration of Ti$^{3+}$ and oxygen ions induced by the off-stoechiometry.

 Powder neutron diffraction patterns were recorded on the diffractometer D1B at the Institut Laue Langevin (ILL) with a neutron wavelength $\lambda$ = 2.52 \AA, down to 0.07\,K. As shown previously \cite{Gardner1999,Mirebeau2002} a typical powder pattern of \tbti\ consists of nuclear Bragg peaks coexisting with a large diffuse signal from the environmental background, incoherent scattering, and short range magnetic  correlations between first neighbors Tb$^{3+}$ moments. These $"$liquid-like$"$ correlations start to appear \cite{Gardner1999,Mirebeau2002} below about 50\,K.  The mesoscopic magnetic order studied here consists of small Lorentzian peaks which appear on the top of this diffuse signal on the pattern measured at 0.07\,K (Fig. \ref{figunsubtracted}). To isolate them from the other contributions, a neutron pattern measured at 1.2(1)\,K was subtracted. The amplitude of the ordered magnetic moment was calibrated to that of the nuclear Bragg peaks by refining the crystal structure on the pattern at 0.07\,K (Inset Fig. \ref{figunsubtracted}).
 
 Specific heat data $C_{\rm p}(T)$ were recorded below 20~K and down to 0.4~K using the relaxation method with a Physical Property Measurement System (Quantum Design, Inc.). The phonon contribution to the total specific heat can be safely neglected below 8~K. In the temperature range of the measurements, i.e.\ above 0.4~K, the nuclear specific heat varies with the temperature as $T^{-2}$. Following an earlier estimate \cite{Yaouanc2011}, its contribution corresponds to 0.35 and 0.155~J/K.mol at 0.4 and 0.6~K, respectively. The influence of the nuclear part in the upturn of the specific heat below 0.5~K for the $x = +0.01$ and $x = 0$ samples is therefore marginal.
 
 \begin{figure}[ht]
 \centerline{\includegraphics[width=8 cm]{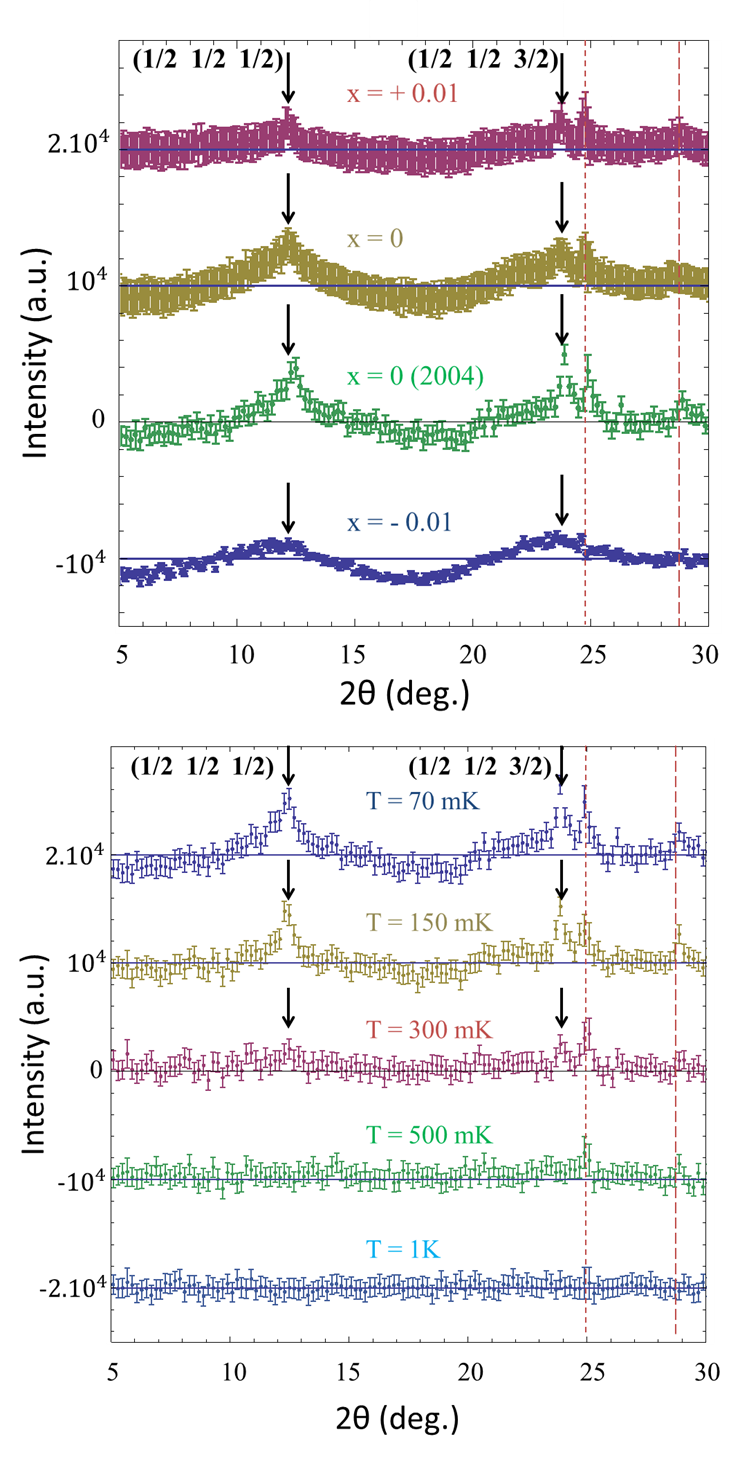}}
\caption{
(color on line) Tb$_{2+x}$Ti$_{2-x}$O$_{7+y}$ : Neutron magnetic patterns, focusing on the region of the most intense magnetic peaks. A pattern at 1.2(1)\,K was subtracted to extract the $\bf{k}$= ($\frac12$, $\frac12$, $\frac12$) mesoscopic order. (a) neutron patterns at 0.07\,K for different Tb contents $x$. Intensities have been scaled to the intensity of a nuclear Bragg peak to allow a comparison between the samples. (b) Neutron patterns versus temperature for the sample $x$=0 (2004).
  Short and long dashed lines shows the positions of (111) and (200) Bragg peaks respectively.
}
\label{figpatterns}
\end{figure}
  
\section{Results}
The lattice constant deduced from X-rays 
increases linearly with $x$. The two stoechiometric samples have close values (a=10.15214(5) and 10.15229(5)\AA\ at 300\,K for the x=0 (2004) and x=0 sample respectively). The values for x=0 and 0.01 are slightly lower than reported in Ref. \onlinecite{Taniguchi2013}, by less than  5 10$^{-4}$\AA\ (Inset Fig. \ref{figcp}a). This difference could be connected with the initial characteristics of the starting constituents, such as the particle sizes of the starting powders, or with some differences in the thermal treatment procedures such as different cooling rates at the end of the thermal treatments and annealing. 

Specific heat data provide a direct evidence of the exotic transition \cite{Yaouanc2011}. 
 As shown in Fig. \ref{figcp}, The $C_{\rm p}(T)$ curves of all samples overlap at high temperature. They strongly differ below 1\,K, showing either a decrease ($x$=-0.01) or an upturn followed by a sharp peak ($x$=0, 0.01). These systematic changes with $x$ reflect the phase diagram of Ref. \onlinecite{Taniguchi2013}. The $x$=-0.01 sample situates in the $"$spin liquid$"$ region of Ref. \onlinecite{Taniguchi2013}. For $x$=0 and $x$=+0.01, a phase transition occurs at 0.4\,K. The sample with nominal value $x$=0 synthesized in 2004 does not show such an upturn, possibly due to different characteristics of the starting constituents or slight differences in the thermal treatment. It confirms that a fine tuning of the off-stoechiometry requires a fully reproducible procedure in all the steps of the synthesis.

 \begin{figure}[ht]
\centerline{\includegraphics[width=8cm]{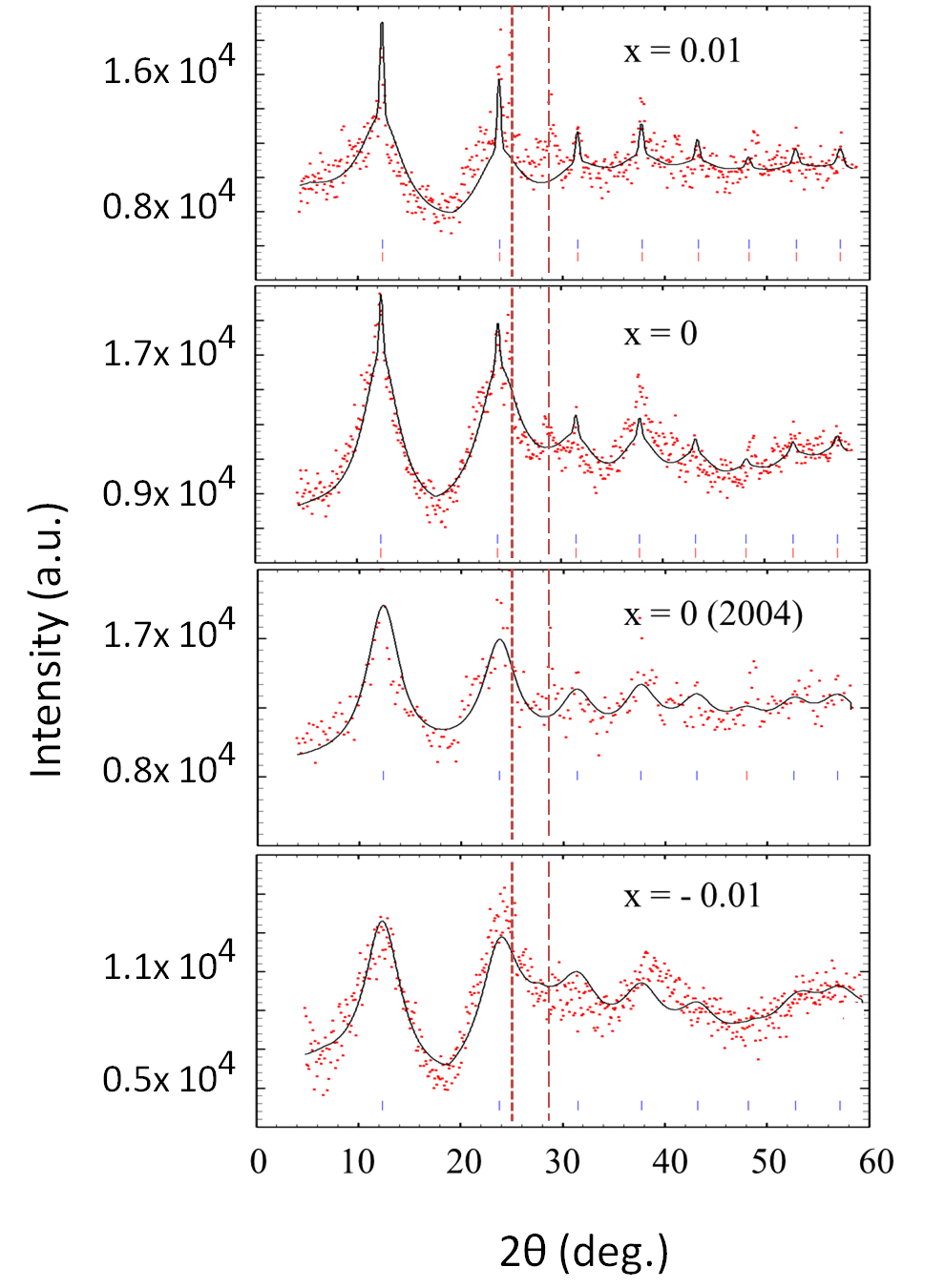}}
\caption{
(color on line) Tb$_{2+x}$Ti$_{2-x}$O$_{7+y}$ : Neutron magnetic patterns at 0.07\,K for different Tb contents $x$. A pattern at 1.2(1)\,K was subtracted. Scales have been enlarged and error bars omitted for clarity. Blue and red tick marks show the positions of the magnetic Bragg peaks for the mesoscopic and long range order respectively. Short and long dashed lines show narrow signals at the position of the (111) and (200) Bragg peaks respectively. Solid lines are the best refinements corresponding to spin ice-like and monopole-like structures, both yielding the same powder averaged intensity (see text).
}
\label{figfitted}
\end{figure}

 The magnetic patterns of all samples are shown in Fig. \ref{figpatterns}(a) at the lowest temperature 0.07\,K, focusing on the low angle region. Intensities were scaled to the intensity of a nuclear Bragg reflection to allow a direct comparison between the samples. The patterns are quite similar, showing broad magnetic peaks with Lorenzian lineshape. In the ($x$=0, 0.01) samples, a small  Bragg-like reflection of width limited by the experimental resolution can be detected on top of the strongest peaks. This Bragg-like contribution is clearly absent in the $x$=-0.01 sample. The Lorenzian peaks are indexed in the Space Group (SG) $Fd\bar{3}m$ with a $\bf{k}$= ($\frac12$, $\frac12$, $\frac12$) propagation vector. The strongest ($\frac12$ $\frac12$ $\frac12$) and ($\frac12$ $\frac12$ $\frac32$) peaks are clearly seen at low angles. The ($\frac12$ $\frac12$ $\frac52$) peaks and weaker peaks with higher Miller indices can be detected at higher angles in the enlarged patterns (Figure 4). The temperature dependence of the magnetic pattern is shown in Figure 3(b) for the sample x=0 (2004). The mesoscopic magnetic order appears below about 0.3\,K.  Small narrow contributions are also detected at the positions of the (111) intense nuclear Bragg peak and at the (200) peak which corresponds to a nuclear extinction. There are respectively attributed to an imperfect data correction and to a minority phase with $\bf{k}$=0 propagation vector similar to that of \tbsn, and not considered in the analysis given below. 

\section{neutron data analysis}

 We first performed a symmetry analysis in the SG $Fd\bar{3}m$, with $\bf{k}$= ($\frac12$, $\frac12$, $\frac12$) using the program BasIreps of the Fullprof suite \cite{Fullprof} and the program Sarah \cite{Wills2000}. As already discussed for \gdti \cite{Champion2001,Stewart2004}, the symmetry operations which keep $\bf{k}$ invariant (symmetry operations of the small Group G$_k$)  split the four rare earth sites of the pyrochlore lattice in two orbits: the site Tb$_1$ with local anisotropy axis along $\bf{k}$ remains invariant (orbit 1), whereas the three other sites  Tb$_{2-4}$ are exchanged (orbit 2). 
 
 The decomposition of the magnetic representation $\Gamma_{\rm Tb}$  for orbits 1 and 2 are: 
 \begin{equation}
{\Gamma_{\rm Tb}^{\rm orbit 1}} = {\Gamma_{\rm 3}^{\rm (1)}}+{\Gamma_{\rm 5}^{\rm (2)}}
\label{orbit1}
\end{equation}
\begin{equation}
{\Gamma_{\rm Tb}^{\rm orbit 2}} = 2{\Gamma_{\rm 2}^{\rm (1)}}+ {\Gamma_{\rm 4}^{\rm (1)}}+ 3{\Gamma_{\rm 6}^{\rm (2)}}
\label{orbit2}
\end{equation}

 We show in Table~\ref{t:IrrRep} the Irreducible Representations (IR) and associated basis vectors (BV) from Sarah output, which provides real basis vectors.  The four sites of a tetrahedron are  Tb$_1$ (0,0,0) for the orbit 1  and Tb$_2$ $(\frac{3}{4},\frac{3}{4},0)$, Tb$_3$ $(0,\frac{3}{4},\frac{3}{4})$ and Tb$_4$  $(\frac{3}{4},0,\frac{3}{4})$, for the orbit 2, which are exchanged by the symmetry operations (xyz), (zxy) and (yzx). This table is similar to that given in Ref. \onlinecite{Champion2001}.
 
 Systematic refinements using the Fullprof suite were performed to the data,  for all possible combinations of the representations within the two orbits. A general state was constructed by combining the basis vectors for a given representation. Good refinements were obtained with $\Gamma_{3}$ and $\Gamma_{6}$ for the orbit 1 and 2 respectively. Taking into account the strong Ising anisotropy of the crystal field, we implemented additional constraints by constraining the Tb moments to lie along their local $<$111$>$ anisotropy axes and to have equal magnitude for the two orbits. With these constraints, the best refinement found from the symmetry analysis corresponds to  the magnetic structure called  " monopole-like " in the following.

 \begin{table}
\caption{\label{t:IrrRep} Irreducible representations $\Gamma$ and associated basis vectors $\psi$ for the space group $Fd\bar{3}m$ with $\bf{k}$= ($\frac12$, $\frac12$, $\frac12$), calculated using the program Sarah \cite{Wills2000}.  The magnetic Tb atoms are at the sites Tb$_1$: $(0,0,0)$, Tb$_2$: $(\frac{3}{4},\frac{3}{4},0)$, Tb$_3$: $(0,\frac{3}{4},\frac{3}{4})$ and Tb$_4$: $(\frac{3}{4},0,\frac{3}{4})$.}
\begin{ruledtabular}\vspace{1ex}
\begin{tabular}{lcccc}
IR                 &Basis Vector &Atom   & BV components                        &\\
                   &        &       &$x$\hfill      $y$\hfill      $z$ &\\[0.5ex]
\colrule
                   &        &       &                                  &\\[-1.5ex]
  Orbit 1                &        &       &                                  &\\
$\Gamma_{3}$  &$\psi_{1}$&Tb$_1$    &1\hfill         1\hfill         1 &\\
                   &        &       &                                  &\\
$\Gamma_{5}$  &$\psi_{2}$&Tb$_1$    &1\hfill      $\bar{1}$\hfill         0 &\\
 &$\psi_{3}$&Tb$_1$    &1\hfill         1\hfill       $\bar{2}$&\\

\colrule
  &        &       &                                  &\\[-1.5ex]
                     Orbit 2                &        &       &                                  &\\
$\Gamma_{2}$  &$\psi_{4}$  &Tb$_2$    &1\hfill         1\hfill         0 &\\
            	   &        &Tb$_3$    &0\hfill 1\hfill 1 &\\
                   &        &Tb$_4$    &1\hfill         0\hfill 1 &\\
                    &        &       &                                  &\\                 
  &$\psi_{5}$  &Tb$_2$    &0\hfill         0\hfill         1 &\\
            	   &        &Tb$_3$    &1\hfill 0\hfill 0 &\\
                   &        &Tb$_4$    &0\hfill         1\hfill 0 &\\
                   &        &       &                                  &\\
                   $\Gamma_{4}$  &$\psi_{6}$  &Tb$_2$    &1\hfill       $\bar{1}$\hfill         0 &\\
            	   &        &Tb$_3$    &0\hfill 1\hfill $\bar{1}$&\\
                   &        &Tb$_4$    &$\bar{1}$\hfill        0 \hfill 1 &\\
%
                   &        &       &                                  &\\
  $\Gamma_{6}$           	   &$\psi_{7}$   &Tb$_2$    &1\hfill         1\hfill         0 &\\
            	   &        &Tb$_3$    &0\hfill         $\bar{1}$\hfill       $\bar{1}$&\\
                   &        &Tb$_4$    &$\bar{1}$\hfill         0\hfill         $\bar{1}$ &\\
                   &        &       &                                  &\\
                            	   &$\psi_{8}$   &Tb$_2$    &0\hfill         0\hfill         1 &\\
            	   &        &Tb$_3$    &$\bar{1}$\hfill         0\hfill         0 &\\
                   &        &Tb$_4$    &0\hfill         $\bar{1}$\hfill         0 &\\
                   &        &       &                                  &\\
                                	   &$\psi_{9}$   &Tb$_2$    &0\hfill         0\hfill         0 &\\
            	   &        &Tb$_3$    &0\hfill        $\bar{1}$\hfill         1 &\\
                   &        &Tb$_4$    &$\bar{1}$\hfill         0\hfill         1 &\\
                   &        &       &                                  &\\
\end{tabular}
\end{ruledtabular}
\end{table}

 \begin{figure*}[t]
\centerline{
\includegraphics[width=18cm]{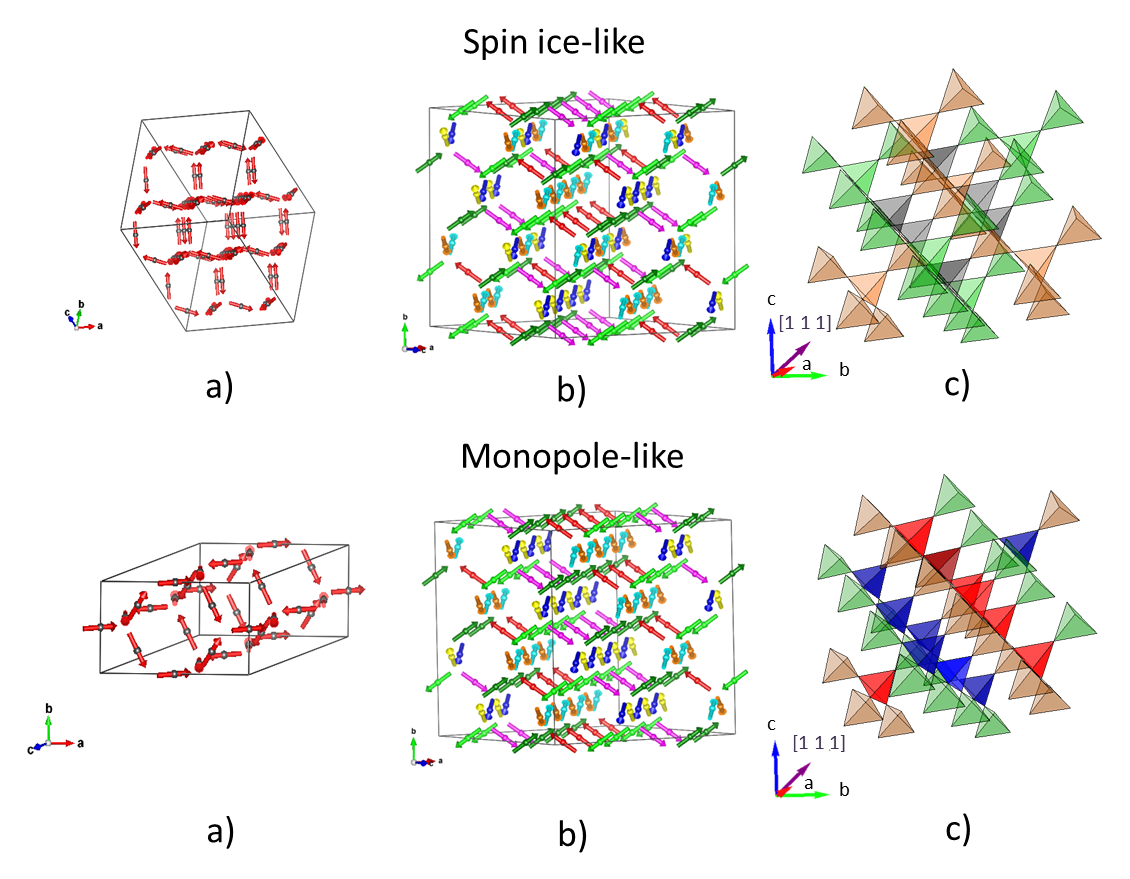}}
\caption{
(color on line) Tb$_{2}$Ti$_{2}$O$_{7}$ : Proposed magnetic structures for the Tb mesoscopic magnetic order. Arrows show the moment orientations. top: spin ice-like. bottom: monopole-like. a): primitive magnetic unit cells. b): cubic magnetic unit cells. Different colors correspond to different spin orientations. c): schematic magnetic stacking of the Tb tetrahedra along a [111] axis parallel to the propagation vector, as shown by a purple arrow. Blue and red tetrahedra correspond to monopole (one-in three-out) and antimonopole (one-out three-in) arrangements respectively. Green and orange tetrahedra correspond to spin ice (two-in two-out) arrangements, with magnetization respectively parallel and antiparallel to the $\it{b}$ axis of the cubic cell. Dark grey tetrahedra correspond to spin ice arrangements with  magnetization along $\it{a}$ or $\it{c}$\ axes of the cubic cell.}
\label{figstructure}
\end{figure*}
  
 We also performed an extensive search in the cubic unit cell, of the magnetic structures  compelling with the above constraints, searching for solutions starting from a Tb tetrahedron. Magnetic structures were refined using the Fullprof Suite \cite{Fullprof}.  We found another magnetic structure, called " spin ice like" in the following, fitting the data equally well as the monopole-like structure. The corresponding refinements are shown in Figure \ref{figpatterns}.  As a striking feature, these two spin structures have  exactly the same powder averaged structure factor. They can be derived from each other by reversing the four moments of appropriate Tb tetrahedra and performing an appropriate translation. Both correspond to complex non collinear antiferromagnetic structures, with cubic magnetic unit cells of length 2$\it{a}$ containing 128 Tb spins, in which the magnetization cancels. Drawings of these two spin structures are shown in Figure \ref{figstructure}. 
  
  In the "spin ice like" magnetic structure (Fig \ref{figstructure} top), the crystal cubic unit cell of length $\it{a}$ contains 4 Tb tetrahedra with the same local ``two-in two-out" spin ice structure. The magnetic cubic unit cell is obtained by translations along the $\langle$100$\rangle$ axes accompanied by spin reversal. This ``spin-ice-like" structure, derived from the ordered spin ice structure 
   of \tbsn \cite{Mirebeau2005} was previously proposed to account for the diffuse ($\frac12$, $\frac12$, $\frac12$) spin correlations in \tbti\ single crystal \cite{Fritsch2013}. It complies with the antiferromagnetic character of the magnetic unit cell by preserving the ice rules through "antiphase" walls in which the local magnetization of a Tb tetrahedron rotates by 90$^\circ$. Relaxing the Ising condition from this structure yields a negligible canting angle (5 deg.) without noticeable improvement of the fit quality. We outline that the "spin ice like" structure cannot be found by a symmetry analysis within the small Group G$_k$, since it violates the face centered cubic symmetry. It should be found  within the decomposition of the General Representation in IR  among the 4 branches of the $\bf{k}$ star, which could also provide multi-k structures. Such decomposition involves IR's up to 12 dimensions, which is beyond the purpose of the paper. We also recall that multi-k structure cannot be distinguished from single-k ones in these powder data.
  
 The "monopole-like" structure (Fig \ref{figstructure} bottom) found by symmetry analysis can be described in the same cubic unit cell as for the spin ice like structure by one monopole ``one-in three-out" and three antimonopole ``one-out three-in" tetrahedra. In other words, it consists of a piling along a $[$111$]$ axis of monopole and antimonopole layers in the Kagome planes, separated by spin ice layers. This ``monopole-like" structure bears some resemblance with the field induced antiferromagnetic one observed in \tbti\ under a high magnetic field oriented along a $[$110$]$ axis, although the propagation vector is different \cite{Sazonov2012}. It complies with the antiferromagnetic character of the magnetic unit cell through a large number of spin ice defects (namely static monopoles), preserving the direction of the staggered magnetization.

 By using the program Isocif on the Bilbao Crystallographic Server \cite{BCS}, we determine the magnetic space group and primitive unit cells (Fig \ref{figstructure}a) of the above structures, which are different from each other. The spin ice-like structure can be generated from a tetragonal magnetic unit cell of unit vectors ${\rm a}_M={\rm b}_M=a\ \sqrt{2},{\rm c}_M=2a$
 containing 64 Tb ions. The primitive magnetic cell of the monopole-like structure is monoclinic ($\beta$ = 125.26$^\circ$, a$_M$ = $a\frac{\sqrt{3}}{\sqrt{2}}$=12.39\AA, b$_M$ = $a\frac{\sqrt{2}}{2}$= 7.15\AA, c$_M$ = $a\sqrt{2}$=14.31\AA) 
 and contains 16 Tb ions. The magnetic space groups are of type IV, listed as follows: (BNS
 I$_c$-4c2 ; OG: P$_I$-4'2m'; Litvin: 111.11.921) and (BNS: C$_c$c ; OG: C$_{2c}$m'; Litvin: 8.6.43) 
 for the spin ice and monopole structures respectively \cite{BCS,Litvin}. They are derived from the Fedorov space groups I-4c2 (N0. 120) and C$_c$ (N0. 9). 
  \begin{table}
\caption{\label{t:moment} Tb$_{2+x}$Ti$_{2-x}$O$_{7+y}$: Mesoscopic magnetic moment $M$, domain size $D$ and long range ordered magnetic moment $M_{\rm LRO}$.}
\begin{ruledtabular}
\begin{tabular}{llrrr}
 $x$ content     &$x=-0.01$ &$x=0(2004)$ &$x=0$   &$x=+0.01$ \\ 
\colrule
$M($\mub$)$      &0.86(1)  &2.17(3)   &1.89(1)  &0.65(1) \\
D($\AA$)       &21(2)    &24(2)     &22(2)     &22(2) \\
$M_{\rm LRO} ($\mub$)$      &0     &0      &0.2(1)   &0.15(3) \\
$C_{\rm p}(T)$ peak      &No     &No     &Yes     &Yes \\
\end{tabular}
\end{ruledtabular}
\end{table}  

 
By fitting the neutron patterns to the above structures, we determined the ordered magnetic moment amplitude and the correlation length relative to the mesoscopic magnetic order for all samples (Table 2). These two quantities are the only free parameters of the magnetic patterns, and they can be determined separately. The amplitude of the ordered moment determines the peak intensities relative to the nuclear Bragg peaks, whereas the domain size controls the peak width. In all samples, the fitted correlation length of the ($\frac12$, $\frac12$, $\frac12$) correlations is around 22(2) \AA, close to the size of the cubic magnetic cell (20.3 \AA). The ordered moment is strongly enhanced for $x$=0, in the region of the critical point $x_C$, then decreases both in the spin liquid and ordered regions. In addition, for the samples showing a peak of $C_{\rm p}(T)$, a minute long range ordered (LRO) moment of about 0.15 \mub\ is detected. This LRO spin component orders over a length scale of about 500 \AA\ close to the resolution limit with the same magnetic structure. In all cases, the ordered moments remain well below the local Tb moment of 5 \mub\ estimated from the crystal field scheme \cite{Gingras2000,Mirebeau2007} at low temperature. This means that in the ground state, a large component of the Tb moment is correlated only over first neighbor distances, both in the spin liquid and in the ordered region of the phase diagram.
  
  \begin{figure}[ht]
\centerline{\includegraphics[width=8cm]{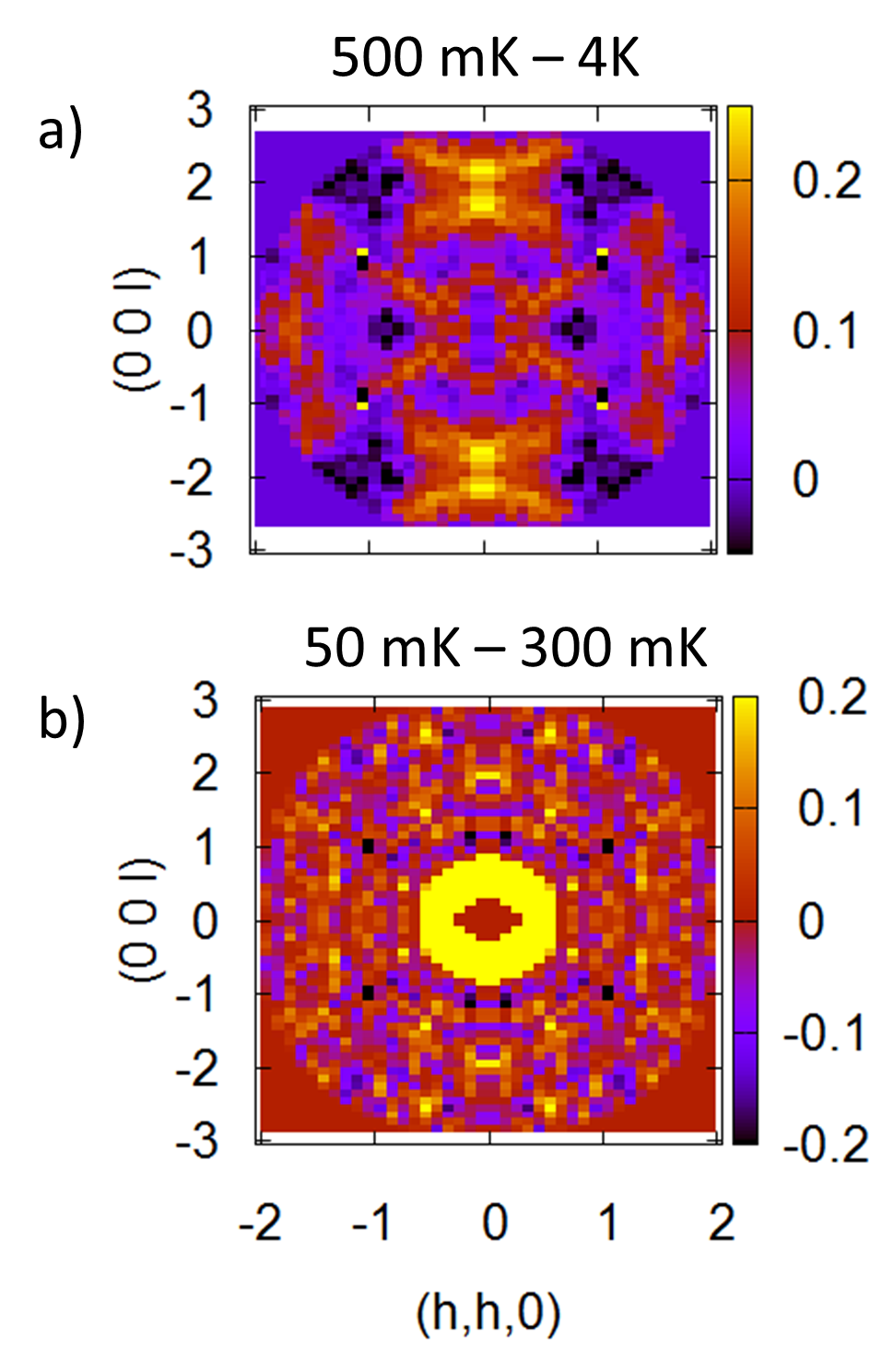}}
\caption{
(color on line) Tb$_{2}$Ti$_{2}$O$_{7}$ single crystal:  Elastic Diffuse scattering maps in a (hh0, 00l) plane. The subtracted maps show the temperature evolution of the correlations. a): between 4\,K and  500 \,mK, the diffuse scattering starts to be structured. b): between 300 and 50\,mK, the correlations do not evolve.}
\label{figmapsubtracted}
\end{figure}

 \section{Discussion}

 The above study focuses on mesoscopic ($\frac12$, $\frac12$, $\frac12$) correlations which emerge in \tbti\ powder patterns below the exotic transition at T$_G$, situated in the range 0.3-0.4\,K,   
  which presents some similarities with a spin glass transition \cite{Yaouanc2011,Lhotel2012,Legl2012,Fritsch2014}. Our analysis provides a phenomenological  quantitative description of these correlations, which involve only a fraction of the Tb moment, amounting to 0.1 to 2 \mub\ depending on the length scale, namely well below the local Tb moment of 5 \mub. These correlations have been separated by subtracting a pattern above the transition. 

  These findings should now be confronted to recent studies in single crystals \cite{Guitteny2013,Fennell2012,Fritsch2014}.  In single crystals, the temperature variation of the diffuse scattering shows a gradual evolution from a fluctuating spin liquid state characterized by isotropic first neighbor correlations at high temperatures, to a precursor regime (between T$_G$ and 4\,K typically) where the diffuse scattering starts to be structured and the elastic component increases steadily, then to a regime (T$<$ T$_G$) where the elastic signal is dominant, well structured and T-independent. This temperature evolution is shown on the subtracted maps of Fig. \ref{figmapsubtracted}.  We speculate that a similar evolution with temperature 
  occurs in powders, so that the mesoscopic correlations can emerge. The precursor regime is not obviously seen here on the powder data due to the low signal background ratio, but it was actually seen in \tbsn\ powder \cite{Mirebeau2007}, where the mesoscopic order occurs on a much larger length scale and involves a much higher fraction of the Tb moment.

 It has been previously argued \cite{Fritsch2014} that there are actually two transitions below 500\,mK, the magnetic transition at T$_G$ occurring about 0.15 \,K below the $C_{\rm p}(T)$ anomaly. This suggests two successive phenomena with decreasing temperature, related to quadrupolar and magnetic order respectively. In the "ordered" samples above x$_c$, the onset of quadrupolar order seen by $C_{\rm p}(T)$  peak occurs at a higher temperature than the long range magnetic order, as frequently observed in rare earth intermetallics \cite{Aleonard1990}. In the "spin liquid" samples below x$_c$ which show no anomaly of $C_{\rm p}(T)$, the magnetic order would remain limited to short length scales.
  
  \begin{figure}[ht]
\centerline{\includegraphics[width=8cm]{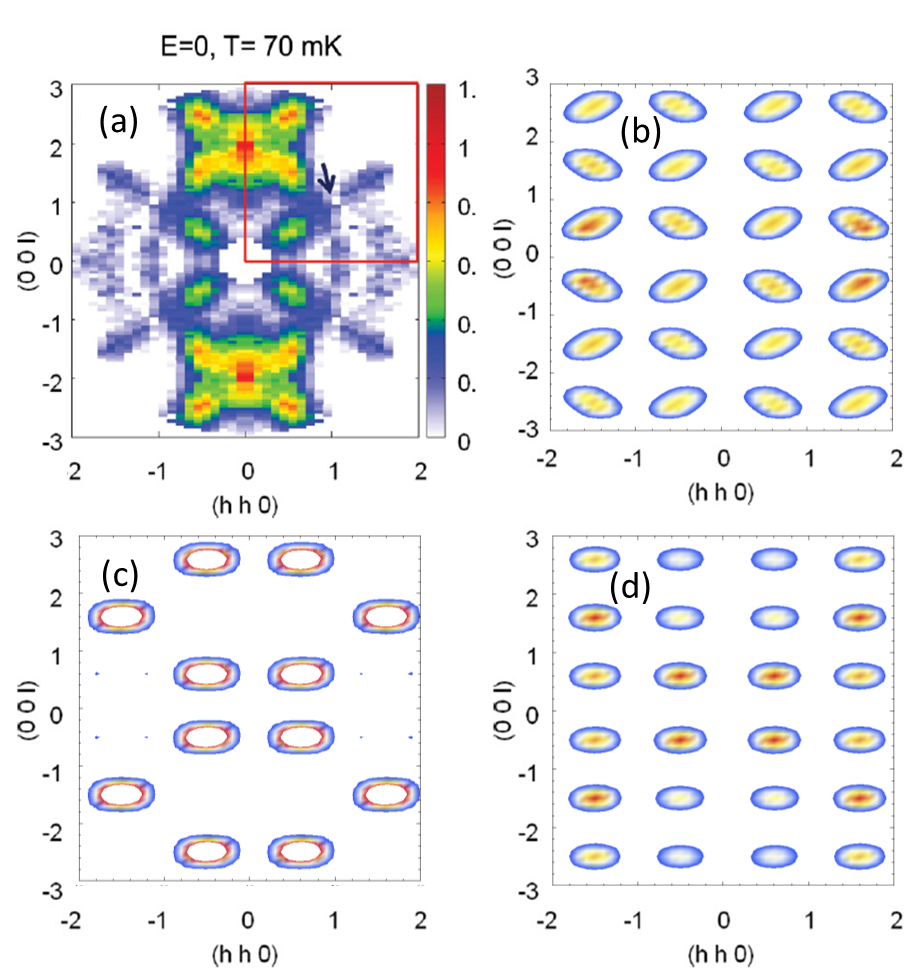}}
\caption{
(color on line) Tb$_{2}$Ti$_{2}$O$_{7}$ : Diffuse scattering maps in a (hh0, 00l) plane at 0.07\,K. (a) : experimental map from Ref. \onlinecite{Guitteny2013}; Calculated maps assuming a single magnetic domain of volume 8$a^3$ with either spin-ice like (b) or monopole-like (c) structure; (d): Average over all single magnetic domains with equal probability. Monopole and spin ice -like structures yield the same average map.}
\label{figmap}
\end{figure}
   
  Below T$_G$ the elastic diffuse scattering measured on single crystals \cite{Guitteny2013,Fennell2012,Fritsch2014} shows arm-like features along $\langle$111$\rangle$ directions and butterfly shaped features centered at (002) and (220). These features result in diffuse maxima around ($\frac12$, $\frac12$, $\frac12$) positions and equivalent (see Fig. \ref{figmap}a).  Moreover, the signal sharpens about certain wave vectors, defining pinch points in reciprocal space. In pyrochlores, these pinch points typically result from the propagation of a local constraint governing the mutual orientations of the magnetic moments \cite{Isakov2004,Henley2005,Moessner1998} such as the "two in/two out"  rule in spin ices. The emergent gauge structure characterizes a Coulomb phase \cite{Henley2005}. Fennell {\it et al} have proposed that a similar ice rule should hold for \tbti, spin components along the local $\langle$111$\rangle$ axes being controlled by a "two in/two out" condition, and in the transverse direction by a "2-up/2-down" condition \cite{Fennell2012}.

   To compare our phenomenological description with these features, we have calculated
    diffuse scattering maps assuming cubic magnetic domains of size 2$\it{a}$ and the spin orientations determined above, for different domain orientations (see the appendix for details). We compare these maps to those measured in the (hh0, 00l) plane by elastic neutron scattering \cite{Guitteny2013} at 0.06(1)\,K shown in Figure \ref{figmap}a. First, we notice that the butterfly features around (002) or (220) as well as the pinch points which appear between the diffuse maxima in the experimental map of Fig \ref{figmap}a are not reproduced by these calculations. Next, for the spin-ice-like structure, a few calculated maps, such as the map of Figure \ref{figmap}b or the maps K1S2 and K1S3 in the Appendix, reproduce rather well the shape, extension and relative orientations of the diffuse maxima at ($\frac12$, $\frac12$, $\frac12$) and related positions. For the monopole-like structure, the agreement is not as good, whatever the domain orientation (see an example in Figure \ref{figmap}c). Averaging randomly over all domains orientations yields the same map for the two structures, shown in Figure \ref{figmap}d, which is also in worse agreement with experiment. This suggests that specific spin-ice like textures, involving magnetic domains of correlated orientations are preferred at the base temperature. They may be generated through nanomagnetic twins, induced by the lowering of symmetry from the  $Fd\bar{3}m$ SG to the tetragonal or monoclinic SG's discussed above. Such nano twins exist in crystal structures \cite{twins} but their extension to magnetism is beyond the purpose of this study.

To situate this picture in the context of a local constraint, we notice that a local constraint supposes the existence of a manifold of degenerate configurations in the ground state. As temperature decreases, the system could select peculiar spin configurations out of this manifold. The spin-ice-like and monopole-like structures determined above would belong to this manifold. This selection could be favored by  small off-stoichiometry such as minute Tb substitution at the Ti sites that would slightly perturb the energy scheme by introducing further neighbor interactions \cite{Cepas2004,Gingras}, local changes in the crystal field, random magneto-elastic couplings or local distortions \cite{Shinaoka2011,Malkin2014}. These perturbations could reveal the basic interactions by inducing correlated regions around impurity sites, with the percolation of small long range ordered moment above a certain level of substitution. Similarly, a low Fe or Co substitution of a Pd matrix reveals the high Pd band susceptibility by generating giant moments around the Fe/Co impurity, with the onset of long range ferromagnetic order below one \% substitution\cite{Low1966}.

One could wonder if such energetic selection supports low energy propagative excitations. This is the case in most fields of physics, where the stabilization of a mesoscopic order with static character allows exotic excitations to emerge. For instance, the slowing down of thermal relaxations allows dispersive excitations in superparamagnetic particles \cite{Hennion1994}, quantum tunneling in molecular magnets \cite{Thomas1996}, reversible magnetization in spin glasses \cite{Alloul1986}, or more trivially spin waves in classical magnets. 
 Here we would expect  excitations to stem from the $\bf{k}$= ($\frac12$, $\frac12$, $\frac12$) positions. Propagative  excitations have been actually observed in \tbti but they stem from the pinch points at $\bf{k}$=(111). This qualitative difference suggests that these excitations are an intrinsic feature of the Coulomb phase.


In summary, we have studied the mesoscopic order at play in Tb$_{2+x}$Ti$_{2-x}$O$_{7+y}$, which is tuned by small off-stoichiometry. We find the same type of correlations in all samples, with two possible structures, monopole-like and spin ice-like, describing the powder averaged magnetic order. Magnetic  moments of about 2 \mub\, are correlated over micro domains of 20 \AA\ size comparable to the cubic magnetic unit cell. Samples in the ordered region of the phase diagram involve in addition a very small long range ordered moment of less that 4\% of the local Tb moment ($\sim$ 5 \mub ). Comparison with single crystal maps suggests that in the ground state, the spin-ice like structures are preferred and that they stabilized through twinned magnetic nano domains. Our observations provide evidence that this mesoscopic order coexists in the ground state with a disordered Coulomb phase, which involves the major part of the local Tb moment. Their links remain however to be elucidated. 
    
We would like to acknowledge fruitful discussions and useful advices from J. Rodriguez-Carvajal, A. Wills, J. Robert, A. Gukasov, F. Damay, and A. Sazonov. We thank M. J. P. Gingras for a critical reading of the manuscript. We are grateful to the ILL for the use of neutron beam time and of the high-flux powder diffractometer D1B operated by the Centre National de la Recherche Scientifique (CNRS). 
   
 \section{Appendix: Magnetic domains}
\begin{figure}[ht]
\centerline{\includegraphics[width=8cm]{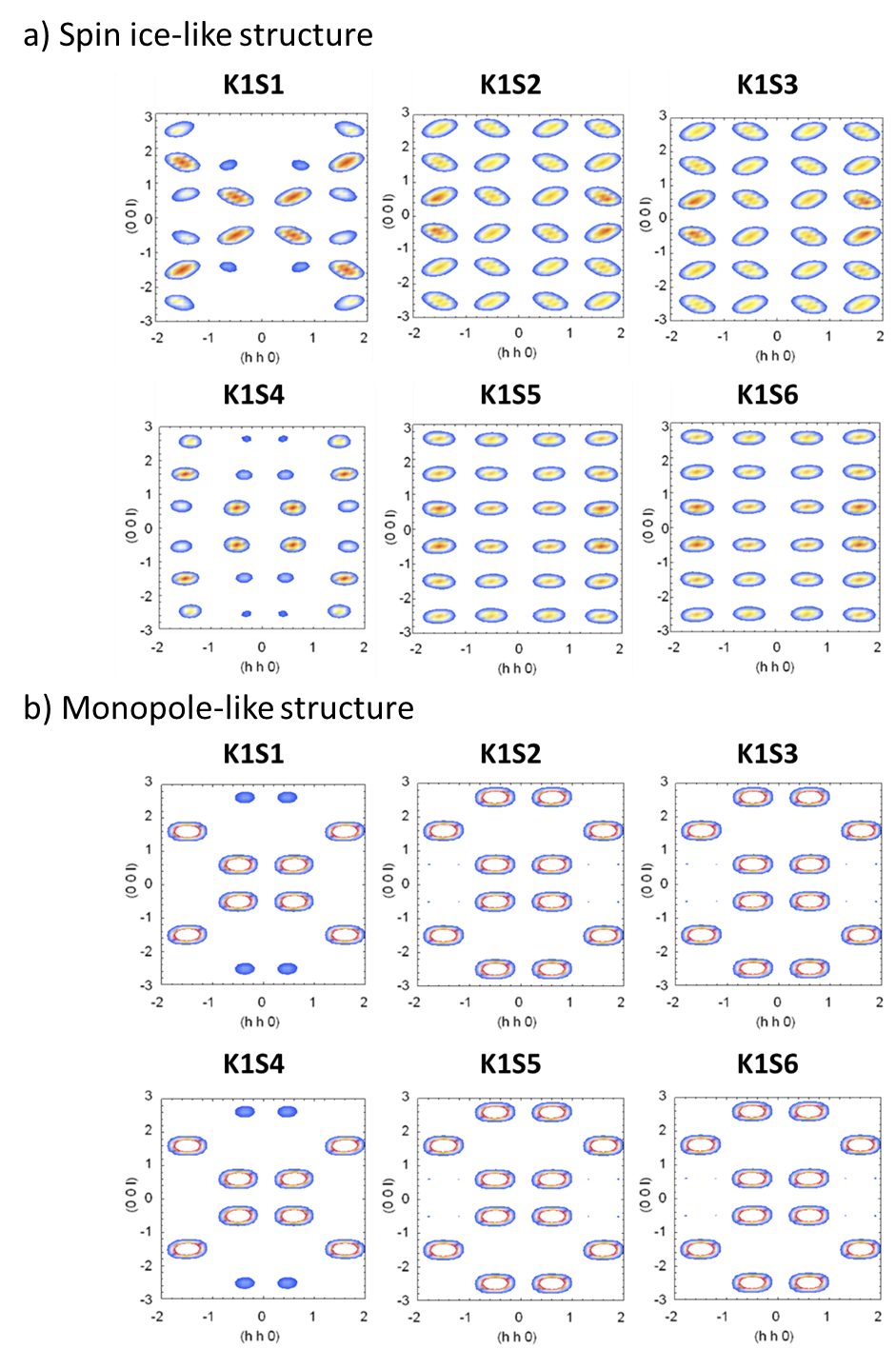}}
\caption{
(color on line) Tb$_{2}$Ti$_{2}$O$_{7}$ : Diffuse scattering maps in the (hh0, 00l) plane at 0.07\,K, calculated for the six S-domains corresponding to the propagation vector $\bf{k}$= ($\frac12$, $\frac12$, $\frac12$). Single magnetic domain of volume 8$\it{a}$$^3$ with spin-ice like (a) and monopole-like (b) structures.}
\label{figdomain}
\end{figure}

 In order to generate the magnetic domains for a given magnetic structure (either spin ice like or monopole like), we used the following procedure. For simplicity, we start from the cubic unit cell of lattice constant $\it{a}$ containing 16 Tb ions. The monopole and spin ice structure are defined within this cubic cell. We generate the K and S-domains within the SG Fd-3m. Among the 48 symmetry operations of the space group Fd-3m, the first 24 are associated with rotation matrices R with det(R)=1, and the 24 others deduced from them by an inversion center (det R=-1) do not need to be considered. Applying the 24 symmetry operations to the propagation vector $\bf{k}$= ($\frac12$, $\frac12$, $\frac12$) splits them in 4 sub groups, which either leave $\bf{k}$ invariant or transform $\bf{k}$ in the $\bf{k}$ vectors (-$\frac12$, $\frac12$, $\frac12$), ($\frac12$, -$\frac12$, $\frac12$), ($\frac12$, $\frac12$, -$\frac12$) of the star \{$\bf{k}$\}. These four subgroups are related to the $\bf{k}$-domains. For the little group G$_k$ which leaves $\bf{k}$ invariant, the symmetry operations apply both to atomic positions and atomic moments, using the usual rules : i) classical spins transform as axial vectors under the symmetry operations; ii) for the symmetry operation which generate atomic positions out of the initial crystal cubic unit cell, equivalent atomic positions/magnetic moments in the cubic cell are obtained by appropriate translations $\bf{R}_{l}$, such as $\bf{M}_{lj}$ = $\bf{M}_{0j}$ exp (2i$\pi$ $\bf{k}$ $\bf{R}_{l}$) where $\bf{M}_{0j}$ and $\bf{M}_{lj}$ are the j moment of the initial cubic unit cell (l=0) and at the distance R$_l$, respectively. This procedure generates for each $\bf{k}$-domain, 6 S-domains, each one corresponding to specific spin orientations of the Tb moments within the cubic unit cell. For each domain, the structure factor and neutron intensity are calculated in the (hh0, 00l) plane, by taking the Fourier transform over a single magnetic cubic cell of size 2$\it{a}$, which corresponds to the typical correlation length deduced from the powder data. We then obtained 24 intensity maps, both for the spin ice-like and monopole-like structure. Figure \ref{figdomain} shows these maps for the six S domains generated with G$_k$, where $\bf{k}$= ($\frac12$, $\frac12$, $\frac12$). In the monopole structure none of the 24 maps corresponds to the single crystal data at low temperature, whereas in the spin ice structure some of them such as the K1S2 and K1S3 shown in Figure \ref{figdomain} indeed bear some resemblance with the data. 
 
  Performing an average over the intensities of the 24 domains with the same weight for each domain, yields the same intensity distribution for the spin ice and monopole structures (Figure \ref{figmap}d) which does not well reproduce the experimental data. It suggests that specific spin textures, associated with  magnetic domains of correlated spin orientations, are actually stabilized. This ground state mesoscopic texture could consist of antiferromagnetically ordered spin ice blocks, with correlated orientations of the staggered magnetization. 
 
\end{document}